\documentclass{article}

    \PassOptionsToPackage{numbers, compress}{natbib}
 \usepackage[preprint]{neurips_2025}

\usepackage{graphicx}
\usepackage{bm} % For \bm command
\usepackage{amsmath}
\usepackage{multirow} % For \multirow command
\usepackage{helvet}  % DO NOT CHANGE THIS
\usepackage{courier}  % DO NOT CHANGE THIS
\usepackage[hyphens]{url}  % DO NOT CHANGE THIS
\usepackage{graphicx} % DO NOT CHANGE THIS
\usepackage[ruled,vlined]{algorithm2e}
\usepackage{natbib}  % DO NOT CHANGE THIS AND DO NOT ADD ANY OPTIONS TO IT
\usepackage{caption} % DO NOT CHANGE THIS AND DO NOT ADD ANY OPTIONS TO IT
\usepackage[utf8]{inputenc} % allow utf-8 input
\usepackage[T1]{fontenc}    % use 8-bit T1 fonts
\usepackage{hyperref}       % hyperlinks
\usepackage{url}            % simple URL typesetting
\usepackage{booktabs}       % professional-quality tables
\usepackage{amsfonts}       % blackboard math symbols
\usepackage{nicefrac}       % compact symbols for 1/2, etc.
\usepackage{microtype}      % microtypography
\usepackage{xcolor}         % colors
\usepackage{enumitem} % 必须加载包
\usepackage{threeparttable}
\usepackage{amsmath,amsfonts,bm}

\newcommand{\mbf}[1]{\mathbf{#1}}

\newcommand{\mcal}[1]{\mathcal{#1}}

\def\eqref#1{equation~\ref{#1}}
\def\1{\bm{1}}

\def\rvu{{\mathbf{i}}}

\def\rvs{{\mathbf{s}}}

\def\rvu{{\mathbf{u}}}

\def\rvx{{\mathbf{x}}}

\def\rvz{{\mathbf{z}}}

\DeclareMathAlphabet{\mathsfit}{\encodingdefault}{\sfdefault}{m}{sl}
\SetMathAlphabet{\mathsfit}{bold}{\encodingdefault}{\sfdefault}{bx}{n}

\def\sR{{\mathbb{R}}}

\usepackage{bm}  

\def\x{{\mathbf{x}}}

\newcommand{\pluseq}{\mathrel{+}=}
\usepackage{siunitx} % for \si\angstorm  Å
\usepackage{amsmath}

\newcommand{\ourM}{BioMD}

\usepackage{caption}  % For caption customization
\newcounter{fig}
\renewcommand{\thefig}{\arabic{fig}} % S1, S2, S3...

\newcommand{\figcaption}[1]{
  \refstepcounter{fig} % 增加计数器的值
  \caption*{\begin{small}\textbf{Figure \thefig.} #1\end{small}} 
}

\newcounter{tab}
\renewcommand{\thetab}{\arabic{tab}} % S1, S2, S3...

\newcommand{\tabcaption}[1]{
  \refstepcounter{tab} % 增加计数器的值
  \caption{\begin{small}\textbf{Table \thetab.} #1\end{small}} 
}

\title{BioMD: All-atom Generative Model for Biomolecular Dynamics Simulation}

\author{
    Bin Feng\thanks{These authors contributed equally to this work.},  Jiying Zhang\footnotemark[1],  Xinni Zhang,  Zijing Liu\thanks{Corresponding author: liuzijing@idea.edu.cn, liyu@idea.edu.cn},  Yu Li\footnotemark[2] \\
  International Digital Economy Academy \\
  Shenzhen, China \\
}

\begin{document}

\maketitle

\begin{abstract}
    Molecular dynamics (MD) simulations are essential tools in computational chemistry and drug discovery, offering crucial insights into dynamic molecular behavior. However, their utility is significantly limited by substantial computational costs, which severely restrict accessible timescales for many biologically relevant processes. Despite the encouraging performance of existing machine learning (ML) methods, they struggle to generate extended biomolecular system trajectories, primarily due to the lack of MD datasets and the large computational demands of modeling long historical trajectories. Here, we introduce BioMD, the first all-atom generative model to simulate long-timescale protein-ligand dynamics using a hierarchical framework of forecasting and interpolation. We demonstrate the effectiveness and versatility of BioMD on the DD-13M (ligand unbinding) and MISATO datasets. For both datasets, BioMD generates highly realistic conformations, showing high physical plausibility and low reconstruction errors. Besides, BioMD successfully generates ligand unbinding paths for 97.1\% of the protein-ligand systems within ten attempts, demonstrating its ability to explore critical unbinding pathways. Collectively, these results establish BioMD as a tool for simulating complex biomolecular processes, offering broad applicability for computational chemistry and drug discovery.
\end{abstract}

\section{Introduction}

Molecular dynamics (MD) simulations have emerged as an indispensable tool in computational chemistry and drug discovery, offering insights into the dynamic behavior of biomolecular systems. Through numerical integration of Newton's equations of motion, MD simulations directly produce atomic trajectories that reveal the time evolution of molecular structures~\citep{hollingsworth2018molecular}. These trajectories enable the exploration of conformational ensembles, optimization of small molecule structures, and identification of potential binding sites, significantly accelerating the design and development of novel therapeutics~\citep{karplus2002molecular}.

Despite their utility, traditional MD simulations face substantial computational limitations. The core bottleneck lies in the intensive calculation of non-bonded forces, particularly van der Waals and electrostatic interactions, which scale quadratically with the number of atoms~\citep{dror2012biomolecular, adcock2006molecular}. Furthermore, accurately resolving high-frequency atomic vibrations necessitates extremely small time steps (on the order of femtoseconds), severely limiting the accessible simulation timescales~\citep{shaw2010atomic, shaw2009millisecond}. Exploring biologically relevant processes, which often span microseconds to milliseconds, remains computationally intensive, restricting the practical application of atomistic MD to obtain trajectories.

\begin{figure}
    \centering
    \includegraphics[width=1.0\linewidth]{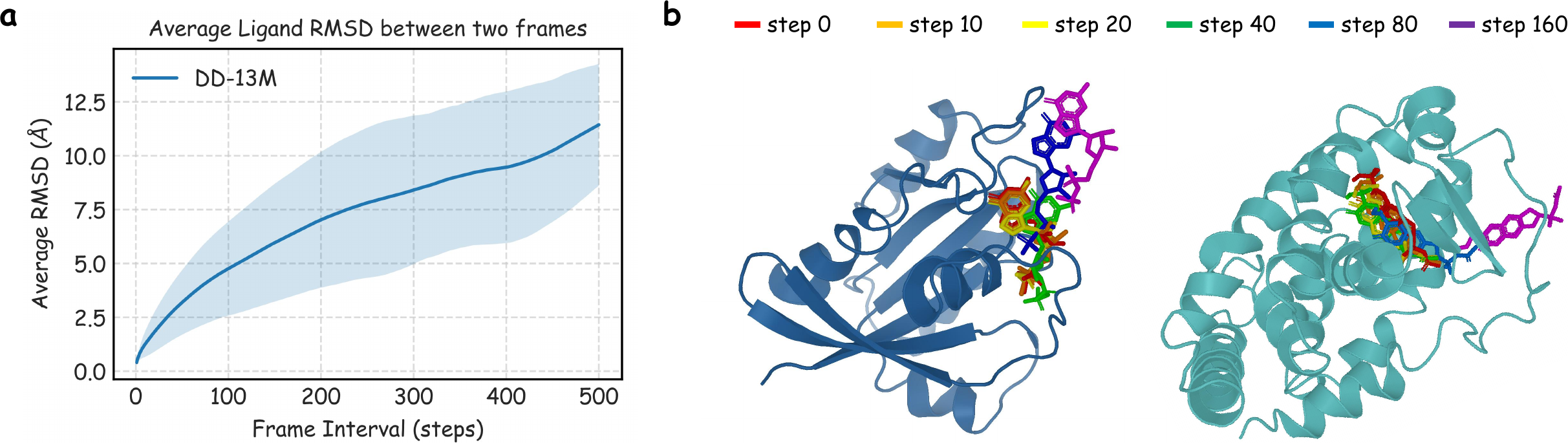}
    \figcaption{\textbf{Average Ligand RMSD between two frames.} (a) Line plot showing that the average ligand RMSD between two frames in the same trajectory increases with the frame interval. (b) Examples of ligand unbinding trajectories at time steps 0, 10, 20, 40, 80, and 160.}
\label{fig:rmsd_example}
\end{figure}

% Recently, ML methods have emerged to overcome these limitations and serve as computational alternatives to MD simulations. Key achievements include protein conformation ensemble generation models~\citep{lewis2025scalable} and highly accurate neural network potentials trained on quantum mechanical calculation data\citep{wang2024ab}. For biomolecular systems, AlphaFold 3~\citep{abramson2024accurate} has shown promising performance for predicting protein-ligand interactions. However, generating MD trajectories for complex protein-ligand systems using purely ML-based approaches remains a significant challenge. This deficiency stems primarily from the inherent complexity of protein-ligand interaction landscapes and the scarcity of high-quality trajectory data for training generative models. There exist some methods for generating biomolecular trajectories. For instance, NeuralMD~\citep{liu2024multi} accelerates protein-ligand binding dynamics simulations by learning trajectories with a neural differential equation solver. However, it neglects the dynamics of protein atoms and considers them static. MDGen~\citep{jing2024generative} utilizes a generative approach to model molecular trajectories for diverse tasks such as forward simulation and transition path sampling. However, it is specially designed for peptides and proteins without considering interactions with small molecular ligands.

Recently, machine learning (ML) methods have emerged as computational alternatives to molecular dynamics (MD) simulations. Key advances include models for generating protein conformation ensembles~\citep{lewis2025scalable} and neural network potentials trained on quantum mechanical data~\citep{wang2024ab}. For biomolecular systems, AlphaFold 3~\citep{abramson2024accurate} has demonstrated promising accuracy in predicting protein--ligand interactions. Despite these achievements, generating full MD trajectories for complex protein--ligand systems using ML remains a major challenge. Existing approaches tend to fall into two categories: (i) methods that can generate protein conformation ensembles but cannot produce time-resolved trajectories~\cite{jingAlphaFoldMeetsFlow2024, wang2024proteinconfdiff}, or (ii) methods that attempt trajectory modeling but struggle to capture protein--ligand interactions. For example, NeuralMD~\citep{liu2024multi} treats protein atoms as static and only models ligand dynamics, while MDGen~\citep{jing2024generative} is specifically designed for peptides and proteins and does not handle small-molecule ligands. This limitation arises from both the complexity of protein-ligand energy landscapes and the scarcity of high-quality trajectory data for training generative models.

To address these limitations, we propose BioMD, a hierarchical framework for generating all-atom biomolecular trajectories. Building upon the insight that short-timescale conformational changes exhibit little conformational change (\textbf{Figure \ref{fig:rmsd_example}}), BioMD decomposes long trajectory generation into two synergistic stages: forecasting of large-step conformations, followed by interpolation to refine intermediate steps. This strategy reduces sequence length by decoupling long-term evolution from local dynamics and helps manage the error accumulation problem for generating long trajectories. Crucially, BioMD unifies forecasting and interpolation within a conditional flow matching model, where we use the ``noising-as-masking'' methods following Diffusion Forcing~\citep{chen2024diffusion} to our time-scale transformer. We apply independent noise to each frame, which enables flexible conditioning on partial trajectory segments, and we implement different tasks simply by using different masking schedules. Inspired by the success of AlphaFold 3, BioMD generates all-atom trajectories using a velocity network that adapts its core transformer architecture, while employing an SE(3)-equivariant graph transformer to encode the initial conformation as conditional embeddings.

To evaluate the effectiveness of BioMD, we conducted experiments on two datasets: MISATO~\citep{siebenmorgen2024misato} and DD-13M~\citep{li2025enhanced}. Our results show that BioMD generates highly realistic conformations with promising physical stability, evidenced by low energy and reconstruction errors across both benchmarks. On the MISATO dataset, which focuses on ligand dynamics within the binding pocket, our model accurately captures the system's conformational flexibility, outperforming existing methods. For the more challenging task of ligand unbinding on the DD-13M dataset, BioMD successfully generates complete unbinding paths for up to 97.1\% of the protein-ligand systems, demonstrating a robust ability to explore critical and long-timescale biomolecular pathways. Collectively, these results establish BioMD as a powerful and efficient tool for simulating complex biomolecular processes, offering broad applicability for computational chemistry and drug discovery.

\section{Related Works}
\paragraph{Conformational Ensemble Generation.}
One major line of research uses ML to generate a biomolecule's conformational ensemble by modeling the equilibrium distribution of its dynamic structures.
Early efforts like AlphaFold2~\cite{jumper2021highly} produce a set of diverse conformations primarily through multiple sequence alignment (MSA) subsampling and masking techniques~\cite{stein2022speach_af, del2022sampling, wayment2024predicting}. 
More advanced approaches now directly learn the conformational distribution from large-scale MD datasets using flow-based~\cite{noe2019boltzmann, jingAlphaFoldMeetsFlow2024} or diffusion-based~\cite{wang2024proteinconfdiff, Eigenfold, str2str,zhengPredictingEquilibriumDistributions2024aDiG,lu2025aligning} generative models. 
Models such as BioEmu~\cite{lewis2025scalable} can effectively generate diverse and physically plausible conformations, providing a powerful alternative to extensive MD sampling to understand a conformational space.
However, these methods are fundamentally time-agnostic; they can sample what conformations are possible but lack the temporal information to show the kinetic pathways between them.

\paragraph{Trajectory Learning for MD Simulation.}
To capture these kinetic pathways, a complementary research direction aims to generate full, time-ordered trajectories.
Approaches like EquiJump~\cite{costa2024equijump} learn to sample future states based solely on the current conformation. 
To capture higher-order dependencies between the frames, MDGen~\cite{jing2024generative} models the joint distribution of entire trajectories via masked frame modeling. 
CONFROVER~\cite{shen2025simultaneous} models these dependencies auto-regressively by conditioning each frame on its entire history through a causal transformer.
While powerful, these methods are often specialized for protein-only dynamics. Conversely, methods that model protein-ligand interactions often introduce other simplifications. For instance, NeuralMD~\cite{liu2024multi} treats the protein receptor as static, which limits the scope of accessible dynamics.

\section{Preliminaries}
\paragraph{Notations.}
A complex $\mathcal{C}$ is composed of a protein $\mathcal{P}$ and a ligand $\ell$. The trajectory of a complex contains $T+1$ frames of coordinates, denoted as $\mbf{X}_T = \{\rvx_{0}, \x_1,\cdots \x_T\} \in \sR^{(T+1)\times N\times 3}$,
% $\{\rvx_{1}^{\mcal{C}}, \x_2^{\mcal{C}},\cdots \x_N^{\mcal{C}}\}$,
where $\rvx_{t}=[\x^\mcal{P}_t, 
\x^{\ell}_t] \in \mathbb{R}^{N \times 3}$ represents the concatenation of protein coordinates $\x^\mcal{P}_t$ and  ligand coordinates $\x^\mcal{\ell}_t$ at time-step $t$, and $N$ is the number of atoms in the complex. The complex trajectory prediction task is defined as generating subsequent conformations (coordinates) of a complex trajectory given its initial conformation (i.e., the first frame).

% \paragraph{Molecular dynamics.} 
% Molecular dynamics (MD) simulates the time evolution of a particle system over \( T > 0 \) using classical mechanics. Each particle follows the relation \( dx_i = p_i / m_i \, d\tau \), where \( \mathbf{p} \in \mathbb{R}^{N \times 3} \) represents the momentum and \( m_i \in \mathbb{R} \) is the mass of particle \( i \). The system's energy is defined by a force field \( E : \mathbb{R}^{N \times 3} \to \mathbb{R} \), and the momentum update follows Newton’s law: \( dp_i = -\nabla_{x_i} E(\mathbf{x}) \, d\tau \). To numerically integrate the dynamics, schemes such as the Verlet algorithm~\citep{Verlet1967} or Langevin dynamics (with a thermostat) are employed. These methods iteratively update positions and momenta, generating a trajectory that approximates the Boltzmann distribution.

\paragraph{Molecular dynamics.} 
Molecular dynamics (MD) simulates the time evolution of a particle system under classical mechanics. 
It leverages numerical schemes such as Verlet integration~\citep{Verlet1967} or Langevin dynamics to generate trajectories approximating the Boltzmann distribution. In the simplest deterministic case with no friction or noise, each particle $i$ evolves according to
$
\mathrm{d}x_i = \frac{p_i}{m_i}\,\mathrm{d} t, 
\mathrm{d}p_i = -\nabla_{x_i} E(x)\,\mathrm{d} t,
$
where $p_i$ and $m_i$ are the momentum and mass, and $E(x)$ is the potential energy function. 
Metadynamics~\citep{laio2002escaping, barducci2011metadynamics, li2025enhanced} extends MD by introducing a history-dependent bias potential $V(s,t)$, constructed over collective variables $s(x)$ as
$
V(s,t) = \sum_{t' < t} w \exp\!\left(-\frac{\|s(x(t)) - s(x(t'))\|^2}{2\sigma^2}\right),
$
where Gaussians of height $w$ and width $\sigma$ are periodically added to discourage revisiting explored states. 
This bias fills free-energy wells and enhances sampling of rare events and transition pathways beyond the reach of standard MD.

\paragraph{Flow matching based models.} 
Flow matching (FM)~\citep{lipman2023flow} is an efficient and simulation-free method for training continuous normalizing flows (CNFs), a class of generative models based on ordinary differential equations (ODEs). In Euclidean space, CNFs define a transformation \( \phi_\tau(\cdot): \mathbb{R}^{N \times 3} \to \mathbb{R}^{N \times 3} \) via an ODE governed by a time-dependent vector field (or velocity) \( v_\tau \):
\begin{align}
\frac{d}{d\tau} \phi_\tau(\mathbf{x}^0) = v_\tau(\phi_\tau(\mathbf{x}^0)), \quad \phi_0(\mathbf{x}^0) = \mathbf{x}^0, \quad \tau\in[0,1],
\end{align}
Here, \( \mathbf{x}^0 \) is sampled from a simple distribution \( p_0 \), and \( \phi_\tau \) evolves it over time $\tau\in[0,1]$ to match the target distribution \( p_1 \) at \( \tau = 1 \). Since \( v_\tau \) is unknown, FM learns \( v_\tau \) by regressing the conditional flow \( u(\phi_\tau(\mathbf{x}^0 | \mathbf{x}^1)) = \frac{d}{d\tau} \phi_\tau(\mathbf{x}^0 | \mathbf{x}^1) \), where \( \phi_\tau(\mathbf{x}^0 | \mathbf{x}^1) \) interpolates between \( \mathbf{x}^0 \sim p_0 \) and \( \mathbf{x}^1 \sim p_1 \).
In our setting, each conformation \( \mathbf{x}_t \in \mathbb{R}^{N \times 3} \) represents a frame in a complex trajectory, and FM is used to generate future frames from an initial structure.

\section{BioMD Method}

\subsection{A Unified Generative Framework via Flow Matching}
\begin{figure*}
    
    \centering
    \includegraphics[width=1.0\linewidth]{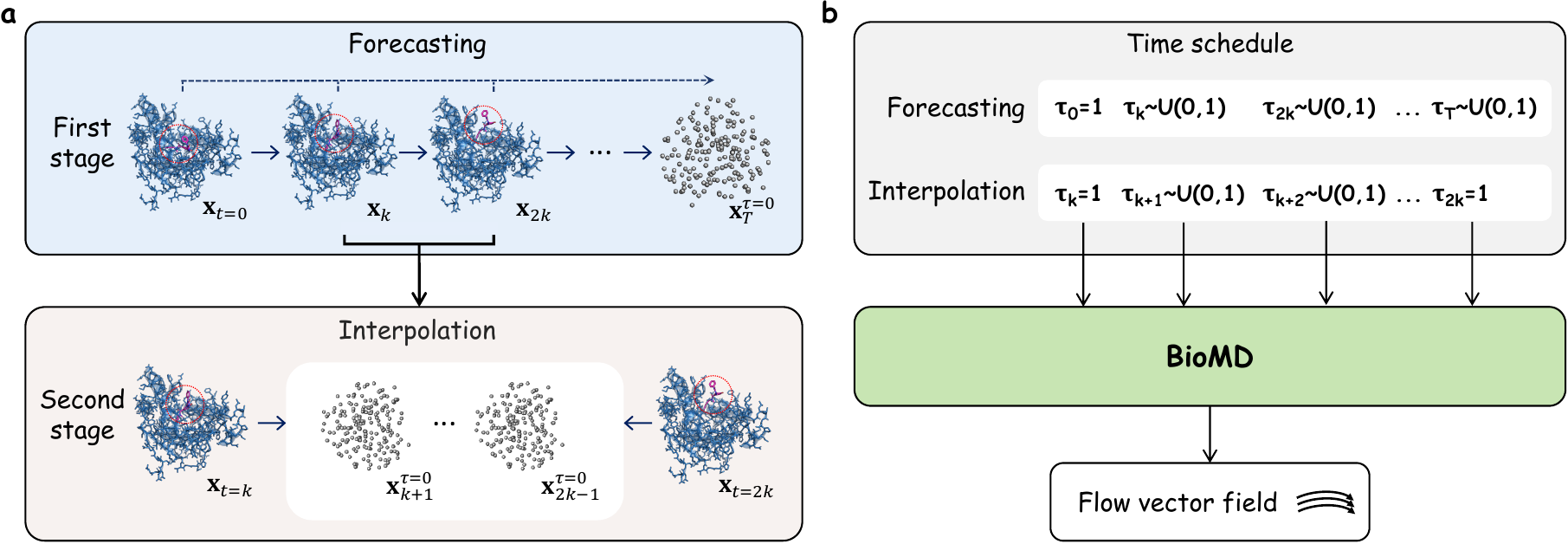}
    \figcaption{\textbf{Model framework.} (a) The hierarchical framework, showing the two-stage process of coarse-grained forecasting followed by fine-grained interpolation. (b) The time scheduling mechanism for forecasting and interpolation tasks, where known frames are noise-free ($\tau=1$) and generated frames are noised ($\tau \in [0,1]$).}
    \label{fig:model_architecture}
\end{figure*}

Our model capitalizes on a fundamental insight into molecular dynamics: conformational changes are typically subtle over short timescales but can involve significant global movements over longer timescales (\textbf{Figure~\ref{fig:rmsd_example}}). This principle underpins our hierarchical prediction framework, which decomposes the generation of long trajectories into two principal stages: coarse-grained forecasting and fine-grained interpolation (\textbf{Figure~\ref{fig:model_architecture}}).

Notably, this entire framework is implemented within a single model architecture that processes the sequence of the whole trajectory at once. We adopt a ``noise as mask'' strategy, where the distinction between the two stages is made simply by varying the input masking patterns (Figure 2b). In this unified framework, each frame in an input sequence is independently perturbed by noise according to a time variable $\tau$. Known or conditioning frames are kept clean (equivalent to setting their corresponding $\tau = 1$, i.e., ``unmasked''), while frames to be generated are initialized from pure noise (equivalent to $\tau = 0$, i.e., ``masked'') and then iteratively denoised.

Let a trajectory sequence be denoted by $\mbf{X} = \{\x_{t_1}, \x_{t_2}, \dots, \x_{t_L}\}$. During training, we sample a vector of independent time steps $\mbf{T} = \{\tau_{t_1}, \tau_{t_2}, \dots, \tau_{t_L}\}$, where each $\tau_{t_i} \sim U(0, 1)$. The sequence is then noised to $\mbf{X}^\mbf{T} = \{\x^{\tau}_{t_1}, \dots, \x^{\tau}_{t_L}\}$, where each frame is an interpolation between the real coordinates and Gaussian noise $\boldsymbol{\epsilon}_i \sim \mathcal{N}(\mbf{0}, \mbf{I})$: $\x^{\tau}_{t_i} = \tau_{t_i} \x_{t_i} + (1-\tau_{t_i})\boldsymbol{\epsilon}_i$. The corresponding ground-truth velocity field for the sequence is $\mbf{U}^\mbf{T} = \{\rvu^{\tau}_{t_1}, \dots, \rvu^{\tau}_{t_L}\}$, with $\rvu^{\tau}_{t_i} = (\x_{t_i} - \x^{\tau}_{t_i}) / (1-\tau_{t_i})$.

Our velocity model $u_\theta$ takes the entire noisy sequence and conditioning information to predict the velocities for all frames simultaneously. The training objective is a Mean Squared Error loss over the entire sequence:
\begin{equation}
    \mcal{L}_{\text{flow}} = \text{MSE}(u_\theta(\mbf{X}^\mbf{T}, \mbf{Z}, \mbf{T}), \mbf{U}^\mbf{T}).
\end{equation}
Here, $\mbf{Z}$ contains static information including the first frame coordinate $\mathbf{x}_0$, amino acid sequence $\mathbf{s}$, and ligand atom types $\mathbf{a}$. We explore two modeling approaches: BioMD-rel, which predicts coordinate changes relative to an anchor frame, and BioMD-abs, which predicts absolute atomic coordinates. For clarity, we focus on the absolute coordinate prediction task below.

\subsection{Hierarchical Generation with Forecasting and Interpolation}
The two stages of our hierarchical framework are realized simply by applying different masking schedules to our unified model during training and inference.

\subsubsection{Coarse-grained Forecasting}
\label{subsec:phase1}
The first stage generates a coarse-grained trajectory, constructed by sampling every $k=10$ steps from the full trajectory, resulting in a sequence $\mbf{X}_C = \{\x_0, \x_k, \x_{2k}, \dots\}$. This task is framed as a forecasting problem where, given the initial frame $\x_0$, the model must generate all subsequent frames.

This is achieved by applying a specific masking schedule to our unified framework. During training, the time step for the initial frame is always fixed at $\tau_0 = 1$ (making it a known, ``unmasked'' condition), while the time steps for all other frames $\{\tau_k, \tau_{2k}, \dots\}$ are sampled independently from $U(0, 1)$. The model $u_\theta$ is trained to predict the velocities for all frames in the sequence, conditioned on the clean initial frame.

During inference, this setup supports multiple generation strategies:
\begin{itemize}
    \item \textbf{All-at-once:} All future frames $\{\x_k, \x_{2k}, \dots\}$ are generated concurrently. We set $\tau_0=1$, initialize all other frames from noise (i.e., their $\tau$ values start at 0), and use an ODE solver like the Euler method to integrate all frames simultaneously to $\tau=1$.
    \item \textbf{Auto-regressive (AR):} Frames are generated in sequential blocks of size $j$. To generate one such block, the model conditions on the previously generated history. This is controlled by the time variable $\tau$: the $\tau$ values for all frames in the history are held constant at 1, making them clean, ``nmasked'' inputs. The $\tau$ values for all $j$ frames within the current target block are then jointly evolved from 0 to 1 by the ODE solver. This process simultaneously denoises all frames in the block from pure noise to their final structures, using the fixed history as context. Once generated, this block is added to the history, and the process is repeated for the next block until the full trajectory is complete.

\end{itemize}

\subsubsection{Fine-grained Interpolation}
\label{subsec:phase2}
After obtaining the coarse-grained trajectory $\{\x_0, \x_k, \x_{2k}, \dots\}$, the second stage replenishes the intermediate frames. This is an interpolation task, where for each coarse interval, we generate the frames $\{\x_{ik+1}, \dots, \x_{(i+1)k-1}\}$ conditioned on the two "anchor" frames, $\x_{ik}$ and $\x_{(i+1)k}$.

This task uses the exact same velocity model $u_\theta$ and training framework, differing only in the data and masking schedule. The input sequence is now a fine-grained segment $\mbf{X}_I = \{\x_{ik}, \x_{ik+1}, \dots, \x_{(i+1)k}\}$. During training, the anchor frames are designated as known by fixing their time steps $\tau_{ik}=1$ and $\tau_{(i+1)k}=1$. The time steps for all intermediate frames are sampled independently from $U(0, 1)$. The model learns to generate the intermediate trajectory conditioned on the start and end conformations.

During inference, this task is always performed in an ``all-at-once'' manner. The anchor frames $\x_{ik}$ and $\x_{(i+1)k}$ are provided as clean inputs (their $\tau=1$), while all intermediate frames are initialized from noise (their $\tau=0$). The model then simultaneously generates all $k-1$ intermediate frames by integrating them to $\tau=1$. This process is described by:
\begin{equation}
    \hat{\mbf{Y}}_{ik}^{\tau+\Delta\tau} = \hat{\mbf{Y}}_{ik}^{\tau} + u_\theta(\hat{\mbf{X}}_I^\mbf{T}, \mbf{Z}_{\text{seq}}, \mbf{T}) \cdot \Delta\tau,
\end{equation}
where $\mbf{Y}_{ik}$ represents the block of intermediate frames, and the velocity predictions are extracted for only those frames. This hierarchical approach allows BioMD to efficiently generate long, physically plausible trajectories.

\subsection{Velocity Model Architecture}
BioMD is a generative model that operates directly on all-atom Cartesian coordinates. In contrast to approaches that rely on internal coordinates such as coarse-grained backbones and torsion angles, our method directly models all atoms, enabling it to capture subtle structural variations that are critical for realistic biomolecular dynamics. The effectiveness of this all-atom modeling strategy has been demonstrated by state-of-the-art biomolecular structure models like AlphaFold3~\citep{abramson2024accurate}. Notably, our unified model architecture is capable of performing both the forecasting and interpolation tasks (subsec. \ref{subsec:phase1} and \ref{subsec:phase2}) within the same framework.

Our velocity model architecture is specifically tailored for generating trajectories from a single initial structure (\textbf{Figure \ref{fig:model_detailed}}). The model first employs an SE(3) Graph Transformer to encode the initial conformation, creating rich single and pair representations. Subsequently, our core generative module, the \texttt{FlowTrajectoryTransformer} (\textbf{Algorithm \ref{alg:diffusion_transformer}}), operates on the entire trajectory sequence. To effectively capture complex biomolecular dynamics, each block of this transformer incorporates two primary attention mechanisms: \texttt{AttentionPairBias} is responsible for modeling intra-frame spatial interactions, while \texttt{TemporalAttention} specifically addresses inter-frame temporal dependencies by focusing on the same atom or token across different time steps. By stacking these two attention mechanisms, the model can simultaneously process spatial and temporal information, which is crucial for accurate trajectory prediction.

\section{Experiments}

We evaluate {\ourM} on two MD trajectory datasets: the MISATO Dataset~\citep{siebenmorgen2024misato}, which comprises protein-ligand interaction trajectories focusing on ligand movement within the protein binding pocket; and the DD-13M Dataset~\citep{li2025enhanced}, which contains trajectories of ligand unbinding from protein binding pockets and ultimately reaching the protein surface. 
Examples of predicted trajectories can be obtained from Zenodo
.~\footnote {\url{https://doi.org/10.5281/zenodo.16979768}}

To comprehensively evaluate our model's performance in generating all-atom biomolecular trajectories, we first evaluate the physical stability of the generated structures. For the DD-13M ligand unbinding dataset, we next evaluate the ligand unbinding success rate and introduce a ligand centroid trajectory similarity metric to assess the accuracy of the predicted unbinding pathways. For the MISATO dataset, given that this dataset provides conformational ensembles, we further evaluate our model's ability to predict the conformational flexibility of both proteins and ligands. In this paper, we compare {\ourM} with several established ML methods, including DenoisingLD~\citep{fu2022simulate,wu2023diffmd,arts2023two}, GNNMD~\citep{fu2022simulate}, VerletMD~\citep{liu2024multi}, and NeuralMD~\citep{liu2024multi}. We also include a Static model as a baseline, where the initial conformation of the system is held constant throughout the entire trajectory.

\begin{table*}[!t]
\centering
\begin{threeparttable}
\scriptsize
\tabcaption{\textbf{Results on the MISATO dataset.} Comparison of all methods on physical stability (first six metrics) and conformational flexibility (last four metrics). Mean values on the test samples are reported.}

\label{tab:misato_results}
\renewcommand{\arraystretch}{1.1}
\setlength{\tabcolsep}{3pt}
\begin{tabular}{lccccccccccc}
\toprule
\multirow{2}{*}{Method} & \multicolumn{2}{c}{Bond Geometry$^\text{a}$} & \multicolumn{2}{c}{Angle Geometry$^\text{a}$} & \multicolumn{2}{c}{Steric Clashes} & \multicolumn{2}{c}{RMSF Correlation$^\text{b}$} & \multicolumn{2}{c}{RMSF Value$^\text{a,c}$} \\
\cmidrule(r){2-3} \cmidrule(r){4-5} \cmidrule(r){6-7} \cmidrule(r){8-9} \cmidrule(r){10-11}
 & MAE & MSE & MAE & MSE & Intra-Lig & Prot-Lig & Ligand & Protein & Ligand (1.211) & Protein (1.002) \\
\midrule
Static        & .0377 & .0023 & .0575 & .0053 & 0 & 0 & - & - & - & - \\
\midrule
DenoisingLD   & $>10^{10}$ & $>10^{27}$ & .1018 & .0431 & .0160 & .0295 & -0.0290 & - & $>10^{12}$ & -  \\
GNNMD         & .2123 & .1032 & .2115 & .1072 & .3626 & .0028 & -0.0103 & - & .2165 & - \\
NeuralMD-ODE & .0483 & .0076 & .0605 & .0086 & .0114 & .0578 & .3405 & - & .3220 & - \\
NeuralMD-SDE & .0483 & .0076 & \textbf{.0604} & .0086 & .0114 & .0578 & .3405 & - & .3220 & - \\
VerletMD      & 19.73 & 1050 & .5847 & .5482 & .1983 & 3.111 & .3356 & - & .3226 & - \\
\midrule
BioMD-rel   & \textbf{.0395} & \textbf{.0026} & .0655 & \textbf{.0075} & \textbf{.0003} & \textbf{.0006} & \textbf{.4861} & .5945 & .5369 & .5177 \\
BioMD-abs    & .0495 & .0155 & .0709 & .0097 & .0019 & .0023 & .4789 & \textbf{.6854} & \textbf{.7023} & \textbf{.6242}\\
\bottomrule
\end{tabular}
\begin{tablenotes}
    \item[a] Bond geometry (bond length) and RMSF values are in angstroms (\AA). Angle geometry (bond angle) is in radians. 
    \item[b] RMSF Correlation is reported using the Pearson correlation coefficient.
    \item[c] RMSF values for reference trajectories are given in parentheses. Values closer to those of the reference indicate better results.
\end{tablenotes}
\end{threeparttable}
\end{table*}

\begin{figure*}[!t]
    \centering
    \includegraphics[width=1.0\linewidth]{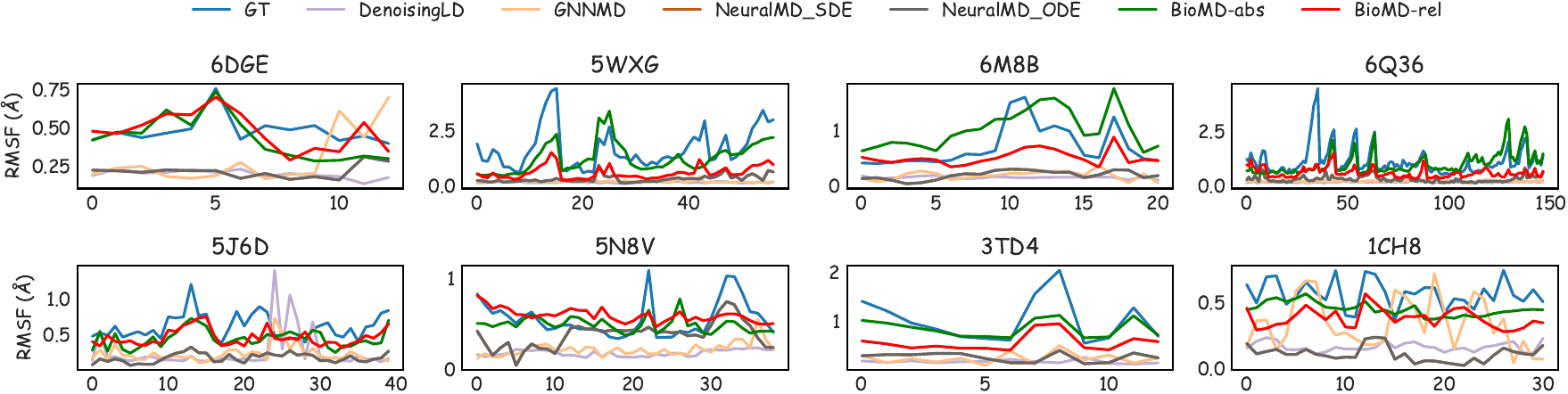}
\figcaption{ \textbf{Ligand RMSF on the MISATO dataset}. Line plot showing Ligand RMSF for eight different protein-ligand systems from the MISATO test set.}
\label{fig:rmsf_plots}
\end{figure*}

\subsection{Results on MISATO}

To evaluate BioMD's ability to generate realistic protein-ligand interaction trajectories, we first conduct experiments on the MISATO dataset, which focuses on ligand dynamics within the protein binding pocket. MISATO comprises nearly 20,000 protein-ligand interaction trajectories, each containing 100 frames sampled from an 8 ns MD simulation. We compare all methods on 1,031 targets with protein sequence length no longer than 800 on the MISATO test set. As shown in \textbf{Table~\ref{tab:misato_results}}, both variants of our model, {\ourM}-rel and {\ourM}-abs, produce trajectories with promising physical stability. The bond and angle geometry errors closely approach the values of the static input structure, and the steric clash scores are orders of magnitude lower than all competing models, confirming the effectiveness of {\ourM} to generate physically plausible structures. 

\begin{figure*}[!t]
    
    \centering
    \includegraphics[width=1.0\linewidth]{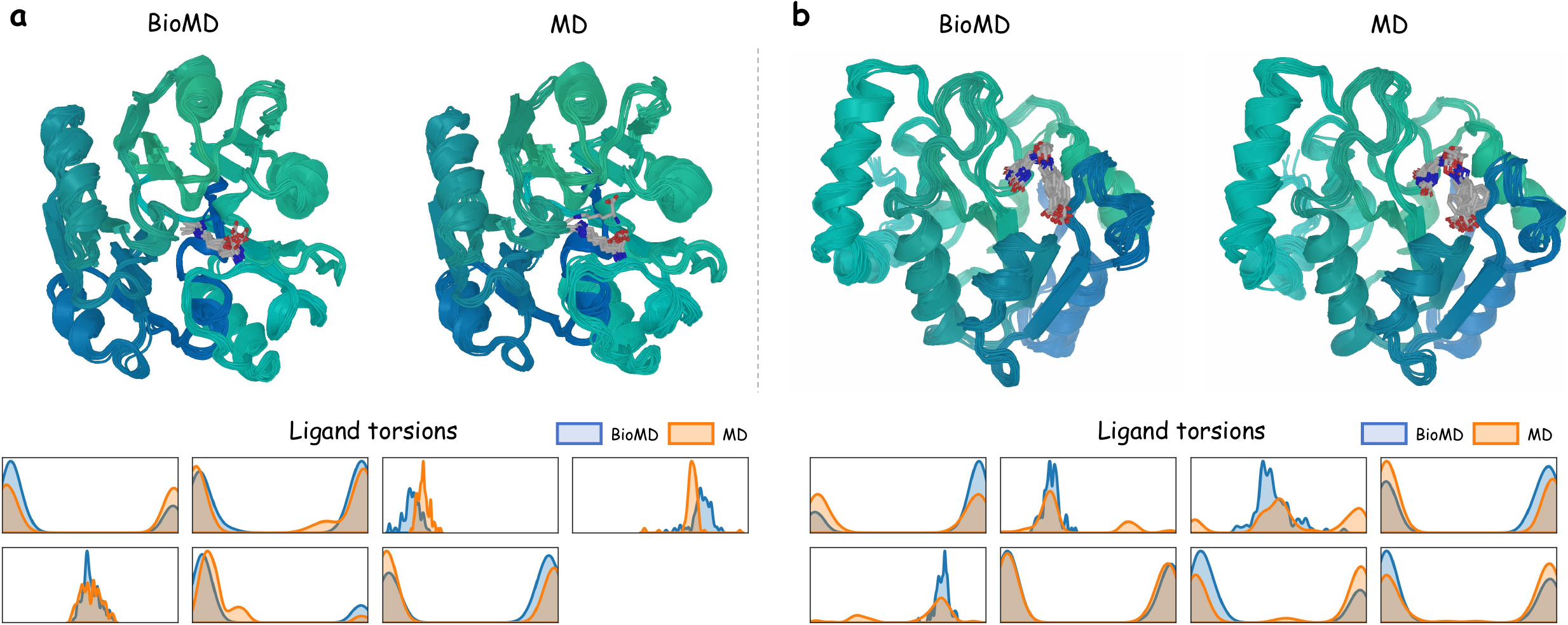}
    \figcaption{\textbf{Conformation ensemble on the MISATO dataset.} A comparison of the distributions of conformations and ligand torsion angles generated by BioMD and MD simulation for 6DGE (\textbf{a}) and 3FCF (\textbf{b}). }
\label{fig:misato_ensemble}
\end{figure*}

In terms of conformational flexibility, BioMD demonstrates a superior ability to capture the system's dynamic behavior. We measure Pearson's correlation between the Root Mean Square Fluctuation (RMSF) of our generated trajectories and the reference MD trajectories. {\ourM} achieves the highest correlation score for ligand atoms, outperforming NeuralMD by 42.8\%. Besides, {\ourM} achieves the correlation score of 0.685 for protein atoms, while other comparing methods fail to simulate protein conformation changes. Visual analysis in \textbf{Figure~\ref{fig:rmsf_plots}} and \textbf{Figure~\ref{fig:misato_ensemble}} further corroborates these findings, showing that BioMD's predicted atomic fluctuations closely trace the ground truth profiles and that the generated conformational ensemble is qualitatively similar to that of a traditional MD simulation. Collectively, these results indicate that {\ourM} can accurately simulate the flexibility of the entire protein-ligand complex.

\subsection{Results on DD-13M}

\begin{figure*}[!t]
    
    \centering
    \includegraphics[width=1.0\linewidth]{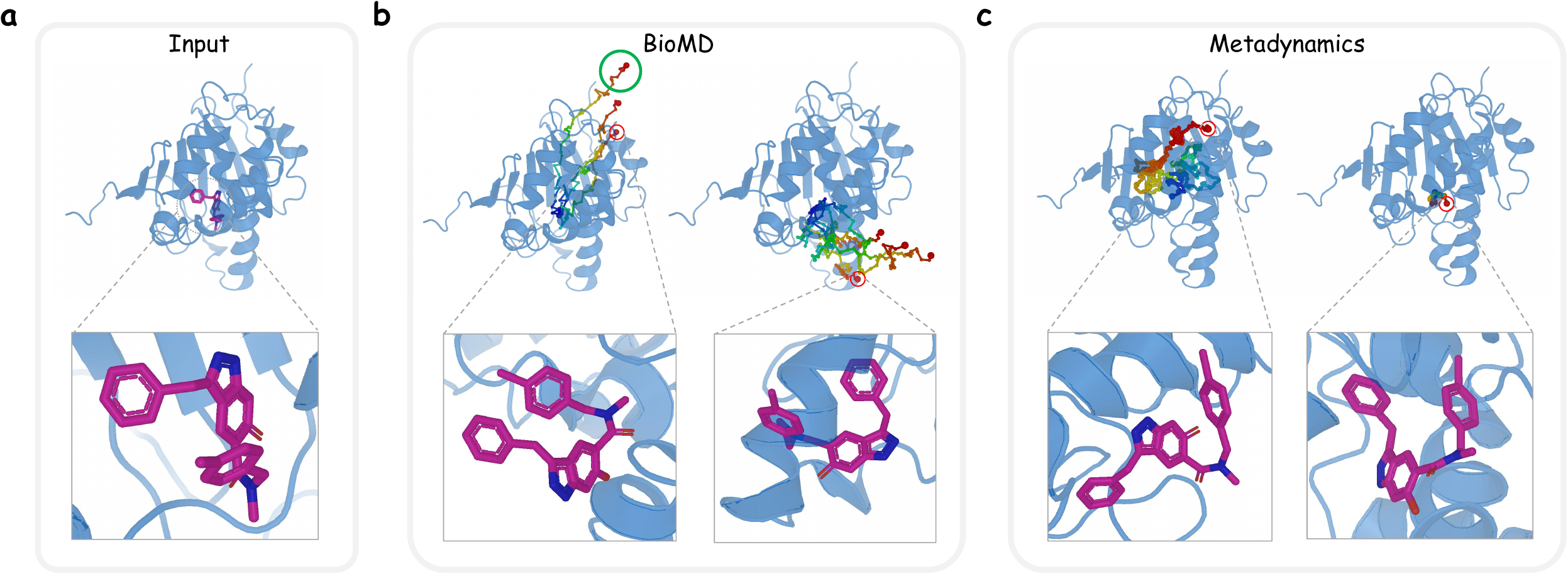}
    \figcaption{\textbf{Ligand unbinding path on 6EY8.} (a) The input conformation. (b) The unbinding pathways generated by BioMD (under 10 seconds), the novel pathway discovered by BioMD is highlighted in a green circle. (c) The reference unbinding pathways obtained using metadynamics simulations (1 hour for the left pathway).}
\label{fig:unbinding_path}

\end{figure*}

We further evaluated BioMD on the more challenging task of ligand unbinding using the DD-13M dataset, which comprises 26,612 dissociation trajectories across 565 complexes, each with an average of 480 frames. 36 complexes were held out as a test set for evaluation, while the remaining were used for training.
A key advantage of our architecture is its flexibility in supporting multiple generation strategies. A concurrent denoising of all future frames, as used on MISATO, results in minimal ligand movement because the model lacks historical guidance and averages over many potential paths. To overcome this, we generate the trajectory auto-regressively, which breaks the long-range prediction into steps and uses previously generated frames to help predict subsequent ones.

The results, summarized in \textbf{Table~\ref{tab:dd13m_results}}, highlight the effectiveness of this auto-regressive strategy. While maintaining high physical stability, the {\ourM}-abs (AR-5) model significantly improved path accuracy, reducing the Unbinding Path RMSD to 0.5645. Most importantly, the AR strategy enabled the successful generation of complete unbinding events. The {\ourM}-rel (AR-5) model achieved a remarkable unbinding success rate, identifying a valid path in 70.9\% of cases with a single attempt (@1), increasing to 92.9\% with five attempts (@5) and 97.1\% with ten attempts (@10). This demonstrates BioMD's reliability in exploring critical biomolecular pathways.

On the qualitative analysis for the 6EY8 system (\textbf{Figure~\ref{fig:unbinding_path}}), our model not only reproduced the two distinct unbinding pathways found by metadynamics simulations with high fidelity but also discovered a novel third pathway, highlighting the exploratory power of our generative approach. Furthermore, BioMD achieves this with remarkable computational efficiency. While metadynamics required 2654 steps (approx. 1 hour on a single GPU) to find the first path, our model generated a complete path in under 10 seconds using just 50 coarse-grained steps.

\begin{table*}[!t]
\centering
\begin{threeparttable}
\tabcaption{\textbf{Results on the DD-13M dataset.} Comparison of methods on physical stability (first six metrics), ligand unbinding path reconstruction metric (Unbinding Path RMSD), and ligand unbinding success rates. Mean values on the test samples are reported.}
\label{tab:dd13m_results}
\scriptsize
\begin{tabular}{lccccccccccc}
\toprule
\multirow{2}{*}{Method} & \multicolumn{2}{c}{Bond Geometry$^\text{a}$} & \multicolumn{2}{c}{Angle Geometry$^\text{a}$} & \multicolumn{2}{c}{Steric Clashes} & {Unbinding Path$^\text{a}$} & \multicolumn{3}{c}{Unbinding Success} \\
\cmidrule(r){2-3} \cmidrule(r){4-5} \cmidrule(r){6-7} \cmidrule(r){8-8} \cmidrule(r){9-11}
 & MAE & MSE & MAE & MSE & Intra-Lig & Prot-Lig & RMSD & @1 & @5 & @10 \\
\midrule
Static & .0254 & .0013 & .0461 & .0037 & .2778 & 0 & .6504 & 0 & 0 & 0 \\ 
Metadynamics$^\text{b}$ & .0246 & .0012 & .0452 & .0030 & .2777 & 0 & .4217 & - & - & - \\ 
\midrule
BioMD-rel & \textbf{.0308} & \textbf{.0018} & .0606 & .0077 & .2943 & .0004 & .6845 & .0029 & .0147 & .0294 \\
BioMD-abs & .0369 & .0026 & \textbf{.0545} & \textbf{.0061} & \textbf{.2941} & \textbf{.0003} & .6802 & .0176 & .0440 & .0588 \\
\midrule
BioMD-rel (AR-5)  & .0580 & .0100 & .0918 & .0184 & .4021 & .6375 & .7055 & \textbf{.7088} & \textbf{.9295} & \textbf{.9706} \\
BioMD-abs (AR-5)  & .0728 & .0111 & .0802 & .0132 & .2943 & .0009 & \textbf{.5645} & .5676 & .7419 & .7941 \\
\bottomrule
\end{tabular}
\begin{tablenotes}
    \item[a] Bond geometry (bond length) and unbinding path RMSD values are in angstroms (\AA), and angle geometry (bond angle) is in radians. 
    \item[b] The metadynamics trajectory serves as the lower-bound. The metrics are calculated among trajectories of multiple repeating simulations.
\end{tablenotes}
\end{threeparttable}
\vspace{-6mm}
\end{table*}

\subsection{Analysis}

The success of the auto-regressive (AR) strategy in modeling long-range dynamics simultaneously exposes a fundamental challenge in generative trajectory modeling: the \textbf{error accumulation problem}. 
As shown in \textbf{Table~\ref{tab:dd13m_results}}, while the non-AR models produce local geometries with errors comparable to the metadynamics reference, the AR models exhibit a notable increase in error. 
However, thanks to our hierarchical framework, these errors remain manageable. The bond and angle MAEs for our AR models remain below 0.1~\AA\ and 0.1~radians, respectively---a threshold well within the range of thermal fluctuations for molecular systems. 
These geometrical errors can be readily corrected via a simple local refinement step with minor structural deviations ($<0.1$~\AA), similar to the relaxation procedure used in AlphaFold.
In contrast, non-hierarchical approaches are trapped between two failure modes: large AR steps yield nearly static trajectories, while small AR steps cause significant error accumulation that results in physically unrealistic structures.

Our results also reveal a distinct trade-off between predicting relative coordinate changes (BioMD-rel) and absolute coordinates (BioMD-abs). The absolute coordinate prediction method (BioMD-abs) demonstrates a superior grasp of the global conformational landscape, evidenced by its higher protein RMSF correlation on MISATO and a more accurate centroid path RMSD on DD-13M, making it the preferred choice for tasks requiring the precise reproduction of specific dynamic pathways. In contrast, the relative coordinate prediction method (BioMD-rel) excels at encouraging more exploratory behavior while preserving local chemical fidelity. Its strength is highlighted by the significantly higher unbinding success rate on DD-13M, which makes it more effective for applications focused on sampling large-scale conformational changes and discovering novel dynamic events. This functional duality means BioMD can be flexibly adapted to the specific goals of a simulation, whether the priority is accuracy in reproducing known dynamics or exploration to discover new ones.

\section{Conclusion}

In this work, we introduce BioMD, a novel all-atom generative model that overcomes the computational limitations of traditional molecular dynamics to simulate long-timescale biomolecular events. Our hierarchical framework, which synergistically combines coarse-grained forecasting with fine-grained interpolation, effectively mitigates error accumulation and enables the generation of physically realistic trajectories. We demonstrated BioMD's capabilities on two challenging datasets, showing it can produce stable conformations that accurately capture protein-ligand flexibility on the MISATO dataset and successfully generate complete ligand unbinding pathways for up to 97.1\% of systems on the DD-13M dataset. Notably, BioMD achieves this with remarkable computational efficiency, identifying unbinding paths in seconds compared to the hours required by traditional methods like metadynamics. By offering distinct modes optimized for either accurate pathway reproduction or broad exploratory sampling, BioMD provides a powerful, flexible, and efficient tool poised to accelerate research in computational chemistry and drug discovery. 

\bibliographystyle{neurips_2025}
\bibliography{ref}

\newpage
\appendix

\section{Technical Appendices and Supplementary Material}\label{app:method_details}

\subsection{Detailed Model Architecture}

\begin{figure*}[!h]
    
    \centering
    \includegraphics[width=1.0\linewidth]{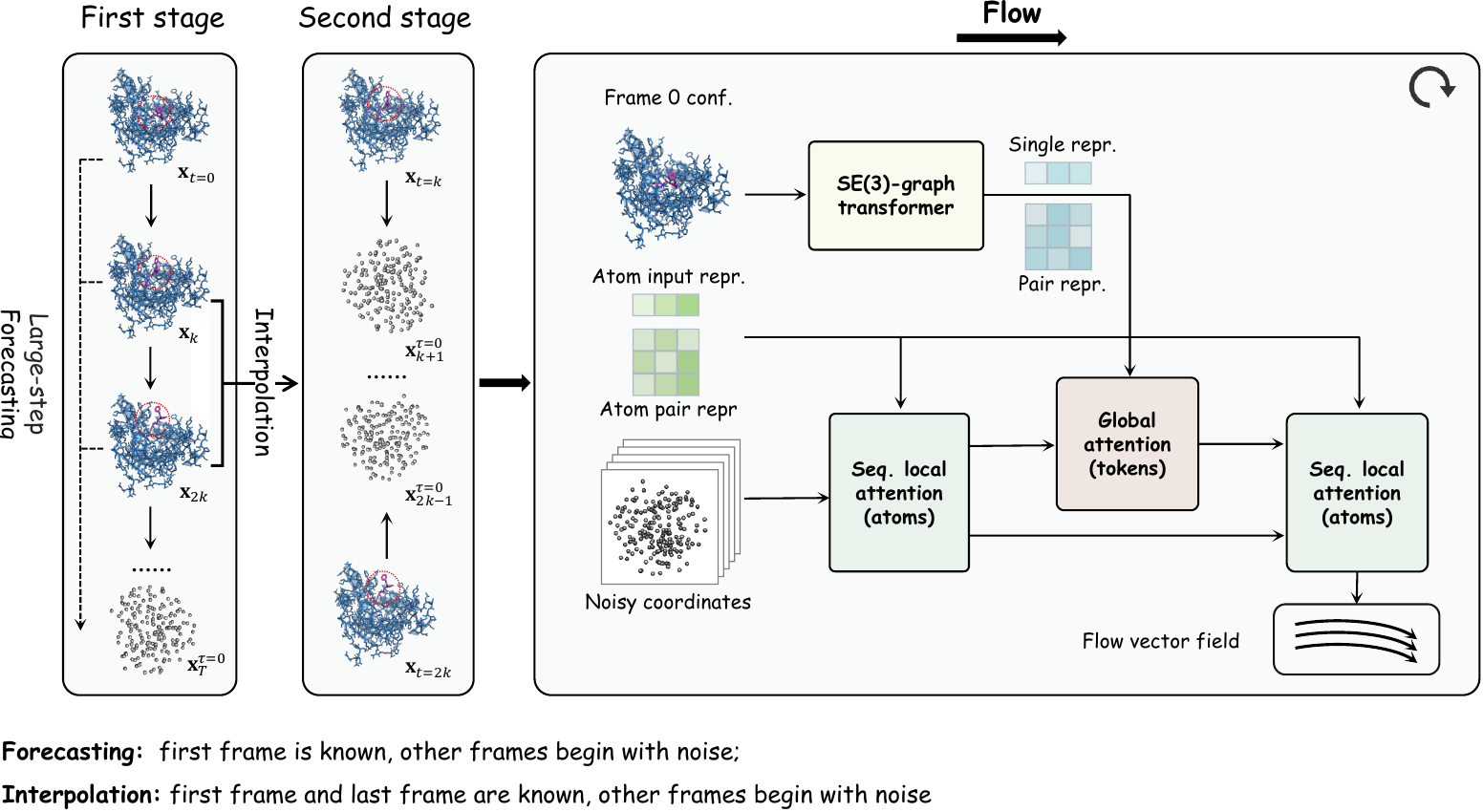}
    \figcaption{\textbf{Detailed architecture of BioMD.} 
The model operates in two modes, \textbf{Forecasting} and \textbf{Interpolation}, set up by the hierarchical framework (left). The core velocity network (right) processes noisy coordinates, conditioned on features from an SE(3)-Graph Transformer. A local-global-local attention pathway generates the final flow vector field used for trajectory generation.}
\label{fig:model_detailed}
\end{figure*}

% (This part is GPT written, need revision)
\paragraph{Hierarchical Generation Framework.}
As illustrated in \textbf{Figure~\ref{fig:model_detailed}}, BioMD employs a hierarchical framework to perform both coarse-grained forecasting and fine-grained interpolation within a unified model. The specific task is controlled by applying noise selectively. For \textbf{Forecasting}, the initial frame $\mathbf{x}_0$ is provided without noise, while all subsequent frames are initialized from a standard Gaussian distribution. For \textbf{Interpolation}, two anchor frames (e.g., $\mathbf{x}_k$ and $\mathbf{x}_{2k}$) are kept clean, while the intermediate frames are initialized from noise. The model's objective is to denoise the masked frames conditioned on the known ones.

\paragraph{Input Representation and Conditioning.}
The core of the model is the \texttt{FlowModule} (\textbf{Algorithm~\ref{alg:flow_module}}), which processes three primary inputs. The main dynamic input is the set of Noisy Coordinates ($\{\vec{\rvx}_l^{\mathrm{noisy}}\}$), representing the current state of the trajectory. To provide structural context, the initial conformation (Frame 0 conf.) is processed by an SE(3)-Graph Transformer, as detailed in the main inference loop (\textbf{Algorithm~\ref{alg:main_inference}}). This produces static Single ($\{\rvs_i^{\mathrm{trunk}}\}$) and Pair ($\{\rvz_{ij}^{\mathrm{trunk}}\}$) representations. These representations, along with other atom features, are processed by the \texttt{FlowConditioning} module (\textbf{Algorithm~\ref{alg:diffusion_conditioning}}) to generate the final conditioning signals.

\paragraph{Spatial-Temporal Attention Pathway.}
The \texttt{FlowModule} uses a local-global-local attention pathway to predict the velocity field. First, the noisy coordinates and conditioning features are passed to an \texttt{AtomAttentionHistoryEncoder}, which models local atomic environments. The resulting representations are aggregated into tokens and fed into the central \texttt{FlowTrajectoryTransformer} (\textbf{Algorithm~\ref{alg:diffusion_transformer}}). This module integrates spatial and temporal information using two key mechanisms: \texttt{AttentionPairBias} resolves intra-frame spatial relationships, while \texttt{TemporalAttention} captures inter-frame dynamics. The globally-aware token representations are then broadcast back to the atomic level, where an \texttt{AtomAttentionDecoder} computes the final per-atom updates.

\paragraph{Velocity Field Prediction and Trajectory Generation.}
The output of the \texttt{FlowModule} is the Flow vector field ($\{\vec{\rvu}_l\}$), which represents the predicted velocity for each atom. During training (\textbf{Algorithm~\ref{alg:train_flow}}), the model is optimized via a mean squared error loss between the predicted velocity and the true velocity. During inference (\textbf{Algorithm~\ref{alg:sample_flow}}), this vector field is used in an Euler integration step, $\vec{\rvx}_l^{\tau+1} \leftarrow \vec{\rvx}_l^{\tau} + dt \cdot \vec{\rvu}_l^\tau$, to iteratively update the coordinates from a noisy state to a final, structured trajectory.

\subsection{Auxiliary losses}
After we get the estimated vector field $\rvu_\theta$, we can get the predicted structure coordinates via
\begin{align}
    \hat{\x}^{1}_t = \hat{\x}^{\tau}_t + \rvu_\theta (1-\tau),
\end{align}
and then we get the predicted protein and ligand structure
$[\hat{\x}^{\mcal{P}}_t,\hat{\x}^{\ell}_t] = \hat{\x}^{1}_t$.

\paragraph{Ligand geometric center loss.}
To stabilize the global placement of the ligand and prevent spurious rigid translations, we align the predicted and reference geometric centers of ligand atoms. Let $\mathbf{x}_t^\ell=\{\mathbf{x}_{t}^{\ell,i}\}_{i=1}^{N_\ell}$ and $\hat{\mathbf{x}}_t^\ell=\{\hat{\mathbf{x}}_{t}^{\ell,i}\}_{i=1}^{N_\ell}$ denote ground-truth and predicted ligand coordinates at step $t$. The geometric center is
\[
C(\mathbf{x}_t^\ell)=\frac{1}{N_\ell}\sum_{i=1}^{N_\ell} \mathbf{x}_{t}^{\ell,i},\qquad
C(\hat{\mathbf{x}}_t^{\ell,i})=\frac{1}{N_\ell}\sum_{i=1}^{N_\ell} \hat{\mathbf{x}}_{t}^{\ell,i},
\]
and the loss is the mean-squared discrepancy
\[
\mathcal{L}_{\mathrm{center}}
= \left\| C(\hat{\mathbf{x}}_t^\ell)-C(\mathbf{x}_t^\ell) \right\|_2^2.
\]
This term softly anchors the ligand’s global position while remaining agnostic to its internal geometry.

\paragraph{Collision loss.}  
To penalize steric clashes, we define a collision loss between protein–ligand atoms and within ligand atoms.  
Let $\x_t^\ell$ and $\x_t^\mcal{P}$ denote ligand and protein atom coordinates at step $t$, and $\hat{\x}_t^\ell$, $\hat{\x}_t^\mcal P$ their predictions.  
We compute predicted distances  
\[
d^{PL}_{ij} = \|\hat{\x}_t^{\mcal P,i} - \hat{\x}_t^{\ell,j}\|_2, 
\quad 
d^{L}_{ij} = \|\hat{\x}_t^{\ell,i} - \hat{\x}_t^{\ell,j}\|_2,
\]
and corresponding ground-truth minimal distances 
\[
d^{PL,gt}_{ij} = \min_t \|\x_t^{\mcal P,i} - \x_t^{\ell,j}\|_2, 
\quad 
d^{LL,gt}_{ij} = \min_t \|\x_t^{\ell,i} - \x_t^{\ell,j}\|_2.
\]

Protein–ligand and ligand–ligand thresholds are set as
\[
\zeta^{PL}_{ij} = \min\!\left(0.9\, d^{PL,gt}_{ij}, \; \zeta_{pl}\right)\, 
\quad
\zeta^{LL}_{ij} = \min\!\left(0.9\, d^{LL,gt}_{ij}, \; \zeta_{ll}\right),
\]
where $\zeta_{pl}=3.0\,$Å and $\zeta_{ll}=2.0\,$Å.  

The collision loss is then defined as
\[
\mathcal{L}_{\text{collision}} = 
\sum_{i,j} \mathbf{1}\!\left(d^{PL}_{ij}<\zeta^{PL}_{ij}\right)\,(\zeta^{PL}_{ij}-d^{PL}_{ij})^2 
+ \sum_{i\neq j} \mathbf{1}\!\left(d^{LL}_{ij}<\zeta^{LL}_{ij}\right)\,(1-b_{ij})\,(\zeta^{LL}_{ij}-d^{LL}_{ij})^2,
\]
where $\mathbf{1}(\cdot)$ represents the indicator function and $b_{ij}$ is the ligand bond mask to exclude bonded pairs.

\paragraph{Ligand bond loss.}  
To preserve ligand bond lengths, we penalize deviations between predicted and ground-truth bonded atom distances.  
Let $\mathcal{B}$ denote the set of bonded atom pairs according to the ligand bond mask.  
For each bond $(i,j) \in \mathcal{B}$, we compute the predicted and ground-truth distances
\[
d^{\ell}_{ij} = \|\hat{\x}_t^{\ell,i} - \hat{\x}_t^{\ell,j}\|_2, 
\quad 
d^{\ell,gt}_{ij} = \|\x_t^{\ell,i} - \x_t^{\ell,j}\|_2.
\]
The bond loss is then defined as the mean squared deviation:
\[
\mathcal{L}_{\text{bond}} 
= \frac{1}{|\mathcal{B}|} \sum_{(i,j)\in\mathcal{B}} 
\left(d^{\ell}_{ij} - d^{\ell,gt}_{ij}\right)^2.
\]

\paragraph{Geometric constraint loss.}
We combine the above terms into a single geometric regularizer
\[
\mathcal{L}_{\mathrm{geom}}
=\lambda_{\mathrm{col}}\,\mathcal{L}_{\mathrm{collision}}
+\lambda_{\mathrm{bond}}\,\mathcal{L}_{\mathrm{bond}}
+\lambda_{\mathrm{ctr}}\,\mathcal{L}_{\mathrm{center}},
\]
where $\lambda_{\mathrm{col}},\lambda_{\mathrm{bond}},\lambda_{\mathrm{ctr}}>0$ balance steric clash avoidance, bond-length preservation, and global ligand anchoring, respectively.

\subsection{Evaluation Metrics}

\subsubsection{Physical Stability}

This metric assesses whether the generated trajectories preserve physically stable conformations, which is essential to ensure chemical validity and avoid unrealistic molecular structures. We evaluate stability from two complementary perspectives:

\begin{enumerate}[leftmargin=*]

\item \textbf{Local Structure Stability.}  
To assess whether the generated trajectories maintain chemically reasonable local geometries, we calculate the deviations of bond lengths and bond angles with respect to the initial frame of the reference trajectories. Both the Mean Absolute Error (MAE) and Mean Squared Error (MSE) are reported. Lower values indicate that the generated conformations remain close to the idealized covalent structure and are thus more chemically stable.

\item \textbf{Steric Clashes.}  
We further quantify the presence of steric conflicts, which occur when non-bonded atoms are unrealistically close to each other. Specifically, a clash is counted if the interatomic distance (excluding bonded pairs and angle-related atoms) is less than a threshold of $1.5$ Å. We compute clash scores for both intra-ligand and protein–ligand interactions, where the score corresponds to the average number of clashes per generated conformation. Lower clash scores indicate physically more plausible conformations.
\end{enumerate}

\subsubsection{Conformational Flexibility}

In addition to stability, it is important that generated trajectories capture the dynamic flexibility of molecular systems. For the MISATO protein–ligand interaction dataset, we adopt the Root Mean Square Fluctuation (RMSF) to quantify the extent of atomic motion over time after trajectory alignment:

\[
\text{RMSF}_i = \sqrt{\frac{1}{T} \sum_{t=1}^{T} \| \mathbf{r}_{i}(t) - \bar{\mathbf{r}}_{i} \|^2},
\]
where $\mathbf{r}_{i}(t)$ is the position of atom $i$ at time $t$, and $\bar{\mathbf{r}}_{i}$ is its time-averaged position.  

We evaluate flexibility from two perspectives:  
1. \textbf{Global Consistency.} We compute the Pearson correlation coefficient between the RMSFs of generated and reference trajectories, where higher correlation indicates better agreement in the fluctuation profiles.  
2. \textbf{Magnitude Accuracy.} We also report the average RMSF of the generated trajectories. Values closer to the reference average RMSF imply that the model produces realistic levels of conformational motion rather than being overly rigid or excessively flexible.

\subsubsection{Unbinding Path Distance}

For the DD-13M ligand unbinding dataset, we evaluate whether generated unbinding trajectories follow realistic spatial pathways compared to reference simulations. We compute the Root Mean Square Deviation (RMSD) between generated and reference ligand centroid trajectories with the following procedure:

\begin{enumerate}[leftmargin=*]
    \item \textbf{Trajectory Standardization.} All trajectories are resampled to a uniform length using linear interpolation, ensuring comparability between different sequences.  
    \item \textbf{Best-Match Search.} For each generated trajectory, we identify the reference trajectory that yields the minimum RMSD. This accounts for the possibility of multiple plausible unbinding pathways.  
    \item \textbf{Final Score.} The reported metric is the average of these best-match RMSDs across all generated trajectories. Lower RMSD values indicate that the model generates ligand motions more consistent with physically realistic unbinding paths.
\end{enumerate}

\subsubsection{Unbinding Success}

This metric evaluates whether the generated ligand trajectories successfully capture the unbinding event. Specifically, we construct the convex hull of the protein heavy atoms in the initial bound state. If at least one predicted ligand centroid position lies outside this convex hull, the trajectory is considered as a successful unbinding case. 

We report the \textbf{Success@k}, which measures the probability that at least one out of $k$ independently generated trajectories for the same protein–ligand complex achieves successful unbinding. A higher success rate indicates a better capability of the model to reproduce realistic ligand unbinding processes. Formally, for each complex with $k$ attempts, Success@k is defined as
\[
\text{Success@}k = \frac{1}{N} \sum_{n=1}^{N} \mathbb{I}\!\left[\max_{1 \leq j \leq k} \; s_{n}^{(j)} = 1\right],
\]
where $s_{n}^{(j)}$ is the binary success indicator (1 if the $j$-th trajectory of complex $n$ achieves unbinding, 0 otherwise), and $N$ is the total number of complexes. We report Success@1, Success@5, and Success@10, which reflect performance under single, moderate, and multiple generation attempts, respectively.

\clearpage

% \subsection{Flow Model Architecture}\label{app:flow_model}

% below, we modify the architectures of $\SampleDiffusion$ and $\DiffusionModule$ from AlphaFold3 \citep{abramson2024accurate} and get the . Notably, we replace the $\mathrm{AtomAttentionEncoder}$ in AF3 with our $\AtomAttentionHistoryEncoder$, which will use our $\FlowTrajectoryTransformer$ instead of the standard $\mathrm{DiffusionTransformer}$.
% (add a history info into Alg. 5 in  and replace the DiffusionTransformer with TemporalSpatial DiffusionTransformer.)

% Elements from these layers are also then incorporated into our custom $InvariantPointAttentionLayer$.
\begin{algorithm}[h]\LinesNumbered
\caption{Main Inference Loop}\label{alg:main_inference}
\KwIn{$\{\mathbf{f}^*\}$ , $\{\vec{\rvx}_{0,l}\}$, $N_{\text{cycle}} = 4$, $c_s = 384$, $c_z = 128$}
$\{\rvs_i^{\mathrm{inputs}}\} \gets \text{InputFeatureEmbedder}(\{\mathbf{f}^*\})$\;
$\rvs_i^{\mathrm{init}} \gets \text{LinearNoBias}(\rvs_i^{\mathrm{inputs}})$\;
$\rvz_{ij}^{\mathrm{init}} \gets \text{LinearNoBias}(\rvs_i^{\mathrm{inputs}}) + \text{LinearNoBias}(\rvs_j^{\mathrm{inputs}})$\;

$\{\rvz_{ij}\}, \{\rvs_i\} \gets 0, 0$\;
\ForEach{$c \in \{1, \ldots, N_{\text{cycle}}\}$}{
    $\rvz_{ij} \gets \rvz_{ij}^{\mathrm{init}} + \text{LinearNoBias}(\text{LayerNorm}(\rvz_{ij}))$\;
        $\{\rvz_{ij}\}, \{\rvs_i\}  \gets \text{\color{blue}GraphTransformer}(\{\vec{\rvx}_{0,l}\},\{\rvs_i\}, \{\rvz_{ij}\}, \{\rvs_i^{\mathrm{inputs}}\})$\;
    $\rvs_i \gets \rvs_i^{\mathrm{init}} + \text{LinearNoBias}(\text{LayerNorm}(\rvs_i))$\;
}
traj\_list=[$\{\vec{\rvx}_{0,l}\}$] \;
\ForEach{$t \in \{1, \ldots, T\}$}{
$\{\vec{\rvx}_{t,l}^{\mathrm{pred}}\} \gets \text{\color{blue}SampleFlow}(\{\vec{\rvx}_{his}\}, \{\mathbf{f}^*\}, \{\rvs_i^{\mathrm{inputs}}\}, \{\rvs_i\}, \{\rvz_{ij}\})$\;
traj\_list.add($\vec{\rvx}_{t}^{\mathrm{pred}}$)\;
$\{\vec{\rvx}_{his}\}$ = traj\_list\;

}

\Return{$\text{traj\_list}$}
\end{algorithm}

\begin{algorithm}[h]\LinesNumbered
\caption{TrainFlow}\label{alg:train_flow}
\KwIn{$\{\vec{\rvx}_l\}$,$\{\vec{\rvx}_{his}\}$, $\{\mathbf{f}^*\}$, $\{\rvs_i^{\mathrm{inputs}}\}$, $\{\rvs_i^{\mathrm{trunk}}\}$, $\{\rvz_{ij}^{\mathrm{trunk}}\}$}
% $\vec{\rvx}_l \sim c_0 \cdot \mathcal{N}(\vec{0}, \mathbf{I}_3)$\;
\textcolor{purple}{\# Indepentent noise levels }\;
$\tau \sim (\mathcal{U}(0, 1),\mathcal{U}(0, 1),\cdots,\mathcal{U}(0, 1))$\;
$\{\vec{\rvx}^0_l\} \sim \mathcal{N}(\vec{0}, \mathbf{I}_3)$\;
% \ForEach{ $\tau$ in $\{0, 0.1,0.2,...,0.9\}$}{
    $\{\vec{\rvx}_l\} \leftarrow \text{CentreRandomAugmentation}(\{\vec{\rvx}_l\})$\;
    % $\gamma \gets \gamma_0$ \textbf{if} $c_\tau > \gamma_{\min}$ \textbf{else} $0$\;
    % $\hat{t} = c_{\tau-1}(\gamma + 1)$\;
    % $\vec{\xi}_l = \lambda \sqrt{\hat{t}^2 - c_{\tau-1}^2} \cdot \mathcal{N}(\vec{0}, \mathbf{I}_3)$\;
    $\{\vec{\rvx}_l^{\tau}\} = 
    \tau \{\vec{\rvx}_l \}+ (1-\tau)\{\vec{\rvx}^0_l\}$\;
    $\{\vec{\rvu}_l^\tau\} \leftarrow \text{\color{blue}FlowModule}(\{\vec{\rvx}_l^{\tau}\}, \{\vec{\rvx}_{his}\}, \tau, \{\mathbf{f}^*\}, \{\rvs_i^{\mathrm{inputs}}\}, \{\rvs_i^{\mathrm{trunk}}\}, \{\rvz_{ij}^{\mathrm{trunk}}\})$\;
    $\mcal{L}_{flow} = \text{MSE}(\{\vec{\rvu}_l^\tau\},\{\frac{\vec{\rvx}_l-\vec{\rvx}_l^\tau}{1-\tau}\} )$\;
    % $\vec{\rvd}_l = (\vec{\rvx}_l - \vec{\rvx}_l^{\text{denoised}}) / {t}$\;
    % $dt = c_\tau - \hat{t}$\;
    % $\vec{\rvx}_l^{\tau+1} \leftarrow \vec{\rvx}_l^{\tau} +  dt \cdot \vec{u}_l^\tau$\;
% }
% \Return{$\{\vec{\rvu}_l^\tau\}$}
\Return{$\mcal{L}_{flow}$}
\end{algorithm}

\begin{algorithm}[h]\LinesNumbered
\caption{SampleFlow}\label{alg:sample_flow}
\KwIn{$\{\vec{\rvx}_{his}\}$, $\{\mathbf{f}^*\}$, $\{\rvs_i^{\mathrm{inputs}}\}$, $\{\rvs_i^{\mathrm{trunk}}\}$, $\{\rvz_{ij}^{\mathrm{trunk}}\}$}

$\vec{\rvx}_l^{0} \sim   \mathcal{N}(\vec{0}, \mathbf{I}_3)$\;
\ForEach{ $\tau$ in $\{0, 0.1,0.2,...,0.9\}$}{
    % $\{\vec{\rvx}_l^{\tau}\} \leftarrow \text{CentreRandomAugmentation}(\{\vec{\rvx}_l\})$\;
    % $\gamma \gets \gamma_0$ \textbf{if} $c\_\tau > \gamma_{\min}$ \textbf{else} $0$\;
    % $\hat{t} = c_{\tau-1}(\gamma + 1)$\;
    % $\vec{\xi}_l = \lambda \sqrt{\hat{t}^2 - c_{\tau-1}^2} \cdot \mathcal{N}(\vec{0}, \mathbf{I}_3)$\;
    % $\vec{\rvx}_l^{\tau} = 
    % \tau \vec{\rvx}_l + (1-\tau)\mathcal{N}(\vec{0}, \mathbf{I}_3)$\;
    $\{\vec{\rvu}_l^\tau\} \leftarrow \text{\color{blue}FlowModule}(\{\vec{\rvx}_l^{\tau}\}, \{\vec{\rvx}_{his}\}, \tau, \{\mathbf{f}^*\}, \{\rvs_i^{\mathrm{inputs}}\}, \{\rvs_i^{\mathrm{trunk}}\}, \{\rvz_{ij}^{\mathrm{trunk}}\})$\;
    
    % $\vec{\rvd}_l = (\vec{\rvx}_l - \vec{\rvx}_l^{\text{denoised}}) / {t}$\;
    % $dt = c_\tau - \hat{t}$\;
    $\vec{\rvx}_l^{\tau+1} \leftarrow \vec{\rvx}_l^{\tau} +  dt \cdot \vec{u}_l^\tau$\;
}
\Return{$\{\vec{\rvx}_l^{1}\}$}
\end{algorithm}

\begin{algorithm}[h]\LinesNumbered
\caption{FlowModule}\label{alg:flow_module}
\KwIn{$\{\vec{\rvx}_l^{\mathrm{noisy}}\}$,$\{\vec{\rvx}_{his}\}$, $t$, $\{\mathbf{f}^*\}$, $\{\rvs_i^{\mathrm{inputs}}\}$, $\{\rvs_i^{\mathrm{trunk}}\}$, $\{\rvz_{ij}^{\mathrm{trunk}}\}$\\
$\sigma_{\mathrm{data}} = 16$, $c_{\mathrm{atom}} = 128$, $c_{\mathrm{atompair}} = 16$, $c_{\mathrm{token}} = 768$}

$\{\rvs_i\}, \{\rvz_{ij}\} \gets \text{\color{blue}FlowConditioning}(t, \{\mathbf{f}^*\}, \{\rvs_i^{\mathrm{inputs}}\}, \{\rvs_i^{\mathrm{trunk}}\}, \{\rvz_{ij}^{\mathrm{trunk}}\}, \sigma_{\mathrm{data}})$\;

% \textcolor{purple}{\# Scale positions to dimensionless vectors with approx. unit variance.}\;
% $\vec{r}_l^{\mathrm{noisy}} = \vec{\rvx}_l^{\mathrm{noisy}} / \sqrt{t^2 + \sigma_{\mathrm{data}}^2}$\;

\textcolor{purple}{\# Sequence-local Atom Attention with history info and aggregation to coarse-grained tokens }\;
$\{a_i\},\{q_k^{\mathrm{skip}}\}, \{p_k^{\mathrm{skip}}\}, \{t_k^{\mathrm{skip}}\} \gets \text{\color{blue} AtomAttentionHistoryEncoder}(\{\vec{\rvx}_{his}\},\{\mathbf{f}^*\}, \{\vec{\rvx}_l^{\mathrm{noisy}}\}, \{\rvs_i\}, \{\rvz_{ij}\}, c_{\mathrm{atom}}, c_{\mathrm{atompair}}, c_{\mathrm{token}})$\;

\textcolor{purple}{\# Full self-attention on token level.}\;
$a_i \gets \text{LinearNoBias}(\text{LayerNorm}(a_i))$\;
$\{a_k\} \gets \text{\color{blue}FlowTrajectoryTransformer}(\{a_i\}, \{\rvs_i\}, \{\rvz_{ij}\}, \beta_{ij} = 0, N_{\mathrm{block}} = 24, N_{\mathrm{head}} = 16)$\;
$a_i \gets \text{LayerNorm}(a_i)$\;

\textcolor{purple}{\# Broadcast token activations to atoms and run Atom Attention.}\;
$\{\vec{\rvu}_l\} \gets \text{AtomAttentionDecoder}(\{a_i\}, \{q_k^{\mathrm{skip}}, p_k^{\mathrm{skip}}, t_k^{\mathrm{skip}}\})$\;

% \textcolor{purple}{\# obtain x1 from vector}\;
% $\vec{\rvx_1} = \vec{\rvx}_l^{\mathrm{noisy}} + (1-t)\vec{\rvu}_l$
% $\vec{\rvx}_l^{\mathrm{out}} = \frac{\sigma_{\mathrm{data}}}{\sqrt{\sigma_{\mathrm{data}}^2 + t^2}} \cdot \vec{\rvx}_l^{\mathrm{noisy}} + \frac{t}{\sqrt{\sigma_{\mathrm{data}}^2 + t^2}} \cdot \vec{r}_l^{\mathrm{update}}$\;

\Return{$\{\vec{\rvu}_l\}$}
\end{algorithm}

\begin{algorithm}[h]\LinesNumbered
\caption{FlowConditioning}\label{alg:diffusion_conditioning}
\KwIn{$\hat{t}$, $\{\mathbf{f}^*\}$, $\{s_i^{\mathrm{inputs}}\}$, $\{s_i^{\mathrm{trunk}}\}$, $\{z_{ij}^{\mathrm{trunk}}\}$, $\sigma_{\mathrm{data}}$, $c_z = 128$, $c_s = 384$}

\textcolor{purple}{\# Pair conditioning}\;
% $z_{ij} \gets \text{concat}([\; z_{ij}^{\mathrm{trunk}},\ \text{RelativePositionEncoding}(\{\mathbf{f}^*\}) \;])$\;
$z_{ij} \gets \text{LinearNoBias}(\text{LayerNorm}(z_{ij}))$\;
\ForEach{$b \in \{1, 2\}$}{
    $z_{ij} \pluseq \text{Transition}(z_{ij},\ n=2)$\;
}

\textcolor{purple}{\# Single conditioning}\;
$s_i \gets \text{concat}([\; s_i^{\mathrm{trunk}},\ s_i^{\mathrm{inputs}} \;])$\;
$s_i \gets \text{LinearNoBias}(\text{LayerNorm}(s_i))$\;
% $n \gets \text{FourierEmbedding}\left(\frac14 \log(\hat{t} / \sigma_{\mathrm{data}}),\ 256\right)$\;
% $s_i \pluseq \text{LinearNoBias}(\text{LayerNorm}(n))$\;
\ForEach{$b \in \{1, 2\}$}{
    $s_i \pluseq \text{Transition}(s_i,\ n=2)$\;
}

\Return{$\{s_i\}, \{z_{ij}\}$}
\end{algorithm}

\begin{algorithm}[h]\LinesNumbered
\caption{FlowTrajectoryTransformer}\label{alg:diffusion_transformer}
\KwIn{$\{a_i\}$, $\{s_i\}$, $\{z_{ij}\}$, $\{\beta_{ij}\}$, $N_{\mathrm{block}}$, $N_{\mathrm{head}}$}

\For{$n \in [1, \ldots, N_{\mathrm{block}}]$}{
    $\{b_i\} \gets \text{AttentionPairBias}(\{a_i\}, \{s_i\}, \{z_{ij}\}, \{\beta_{ij}\}, N_{\mathrm{head}})$\;
    $\{b_i\} \gets \text{\color{blue}TemporalAttention}(\{a_i + b_i\})$\;
    $a_i \gets b_i + \text{ConditionedTransitionBlock}(a_i, s_i)$\;
}

\Return{$\{a_i\}$}
\end{algorithm}

\end{document}